\documentclass{iau}
\usepackage{graphicx,natbib}
 
\newcommand{\apj}{ApJ}           
\newcommand{\apjl}{ApJ}           
\newcommand{\mnras}{MNRAS}       

\newcommand{\aap}{A\&A}

\newcommand{\aj}{AJ}

\newcommand{\apjs}{ApJS}           

\title{Galaxy Mass Assembly with VLT \& HST and lessons for E-ELT/MOSAIC}

\author[Hammer, Flores, Puech]{Fran\c cois Hammer$^1$, Hector Flores$^1$ \& Mathieu Puech$^1$}

\affiliation{
$^1$GEPI, Observatoire de Paris \\ 5 Place Jules Janssen
F-92 195 Meudon, France\\ email: {\tt francois.hammer@obspm.fr}}

\pubyear{2014}
\volume{309}
\jname{Galaxies in 3D across the Universe}
\editors{B. L. Ziegler, F. Combes, H. Dannerbauer, M. Verdugo, eds.}

\begin{document}

\maketitle

\begin{abstract}
The fraction of distant disks and mergers is still debated, while 3D-spectroscopy is revolutionizing the field. However its limited spatial resolution imposes a complimentary HST imagery and a robust analysis procedure.  When applied to observations of IMAGES galaxies at z=0.4-0.8, it reveals that half of the spiral progenitors were in a merger phase, 6 billion year ago. The excellent correspondence between methodologically-based classifications of morphologies and kinematics definitively probes a violent origin of disk galaxies as proposed by \cite{Hammer05}. Examination of nearby galaxy outskirts reveals fossil imprints of such ancient merger events, under the form of well organized stellar streams. Perhaps our neighbor, M31, is the best illustration of an ancient merger, which modeling in 2010 leads to predict the gigantic plane of satellites discovered by \cite{Ibata13}. There are still a lot of discoveries to be done until the ELT era, which will open an avenue for detailed and accurate 3D-spectroscopy of galaxies from the earliest epochs to the present.

\end{abstract}

\firstsection
\section{Methodology to probe distant galaxy kinematics}

 Galaxies assemble their mass through mergers and gas accretion but the balance between these mechanisms is still in debate. 
 Spatially resolved motions are currently observed \citep{Flores06,Forster-Schreiber06,Contini12} at 3 to 7 kpc scales, for galaxies up to z=2.5. Imagery with the HST-ACS reveals sub-kpc details, and galaxy spectral energy distribution is currently retrieved from UV to near-IR (Spitzer) and sub-mm (ALMA) wavelengths. 
 However observations lead to different, sometimes contradictory results \citep{Glazebrook13}, which could require to investigate further, e.g., the various methodologies for distinguishing a disk from a merger.
 
 The first difficulty in 3D spectroscopy, especially at z$>$ 1\footnote{Selecting z$>$1 galaxies using visible filters strongly biases the selection to actively forming, young and low-mass galaxies. Moreover, linking z$>$1 galaxies to their descendants is problematic given the possibility of merger events during the elapsed time, i.e., $³$ 8 billion years.} is to gather complete, mass-selected samples of galaxies. 
Secondly, distant galaxies have small sizes, with stellar half-light diameters decreasing from 2 to 0.6 $arc second$ from z=0.55 to z=2 (for stellar masses larger than $10^{10}MÐ{\odot}$, see \citealt{Dahlen07}). Fortunately ionized gas is currently more extended than the continuum (see Figure 4 of \citealt{Puech10}) and the [OII] emission diameters at z=0.65 average to 2.8 $\pm$0.8 arc second that is covered by $\sim$ 3 spatial resolution elements for a 0.8-1 arc sec FWHM PSF.
At z=2, $H\alpha$ emissions are on average, resolved by the same number of elements: the smaller galaxy sizes are compensated by the better seeing (0.4-0.5 arc second FWHM) in K band\footnote{Observations using adaptive optics are useful to zoom in bright regions in the galaxy cores (e.g., to probe AGN or central outflows, see \citealt{Genzel14, Forster-Schreiber14}), but the outskirts observed with very small spaxels are affected by very low S/N and surface-brightness dimming.}, and the expected larger ionized gas to continuum size ratio. 
 The determination of rotation curves requires more than 7 resolution elements per optical radius (and better 10, e.g., \citealt{Bosma78}), which can be reached by only few massive and extended distant galaxies, or later by the future ELT 3D spectrographs such as MOSAIC.  

With such conditions, interpreting kinematics may appear to be complex though it remains the unique technique for differentiating, e.g., rotating disks from mergers, while smaller perturbations (lopsidedness, warps and bars) are hardly or not detected \citep{Shapiro08}. One needs to elaborate a comprehensive, efficient methodology. Rotating disks possess remarkable geometrical properties, which render them very easy to recognize. Mergers can lead to strong misalignments between (ionized) gas and old star distributions, rotating disks do not. Analysis of kinematics requires a priori a good knowledge of the mass center, the PA and the inclination of a supposed disk \citep{Krajnovic06}. Complimentary deep space-based imagery at red rest-frame wavelengths is thus mandatory to recover the above properties and then to interpret distant galaxy kinematics. The classification procedure for a rotating disk requires to verify, within the observational uncertainties, whether or not:
\begin{enumerate}
\item the mass center coincides to the velocity field center;
\item the difference in PA between red imagery and velocity field is lower than 20 degrees (see \citealt{Epinat12} and references therein and \citealt{Garcia-Lorenzo14});
\item the 'kinemetry' (see \citealt{Shapiro08}) or the '$\sigma$ centering' (see \citealt{Flores06}) are consistent with a dominant rotation motion.
\end{enumerate}
The three steps are mandatory to be verified, and they require mass center, PA and inclination, respectively. The 'kinemetry' method uses an harmonic decomposition of the kinematical maps based on asymmetries. The '$\sigma$ centering' method uses the beam smearing effect: the large and sharp gradient between extremal velocities ($\sim 2 \times V_{max}$) is almost unresolved, and when convolved with random motions, the $\sigma$ map  unavoidably shows a prominent peak at the center of a rotating disk. We notice that 'kinemetry' can be applied on only a tiny fraction of z$\sim$ 2 galaxies (15 among 62 galaxies, \citealt{Forster-Schreiber09}) and cannot distinguish turbulent disks (as those simulated by \citealt{Bournaud10}) from mergers (see \citealt{Perret14}). 

\begin{figure}
\centering
\includegraphics[width=1\columnwidth]{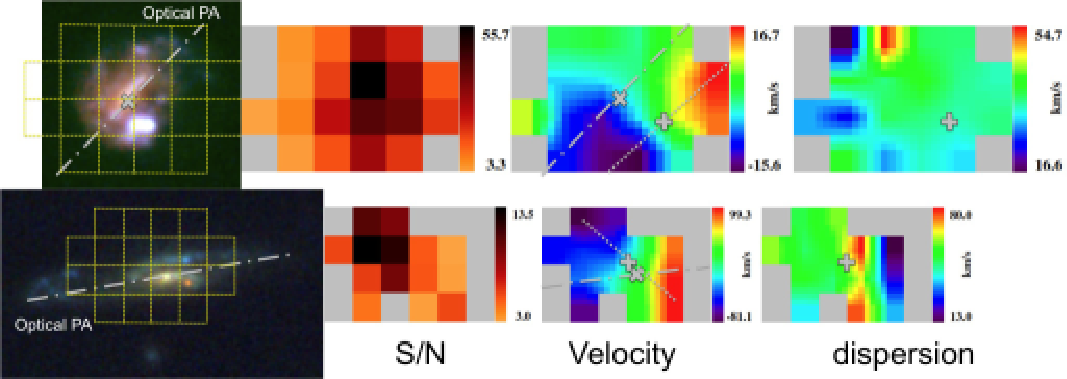} 
\caption{IMAGES galaxies J033239.72-275154.7 at z=0.4151 (top) and J033213.06-274204.8 at z=0.4215 (bottom), with the scale provided by the size of one IFU micro lens, 0.5 arc sec. From left to right: rbg colors using v, i and z filter from HST/ACS, S/N, velocity field and velocity dispersion maps. $PA_{opt}$ and $PA_{VF}$ are indicated by dot-dash and dot lines, respectively. Cross and plus signs mark the optical center ($c_{opt}$) and the velocity field center ($c_{VF}$), allowing to apply the morpho-kinematic classification procedure.}
\label{fig:fig1}
\end{figure}

Figure \ref{fig:fig1} illustrates the three above criteria with two galaxies from IMAGES (see e.g., \citealt{Yang08}). In the top panels, stellar mass and velocity field centers are off by 0.8 arc second, which is much larger than the 0.35 arc second uncertainty, obtained from the quadratic convolution of the astrometry (0.25 arc second) and the line peak location (0.25 arc second) precisions. Interestingly  $PA_{VF}$ is almost aligned with $PA_{opt}$, and without imagery, the system could have been confusedly classified as a rotating disk. However this system would have not passed the "$\sigma$ centering" test, as shown in the top-left panel of Figure \ref{fig:fig1} (see the $\sigma$ peak far offset from the velocity field center). In fact this system has been fully analyzed and modeled as a 3:1, near completion merger by \cite{Peirani09} explaining the giant star-bursting bar, the offset center and the small velocity gradient for a 35 degree inclined galaxy. Figure \ref{fig:fig1} (bottom) shows a galaxy for which morphology resembles well that of an almost edge-on, well defined spiral galaxy. Kinematics is essential here: after examining the S/N map, the most illuminated spaxels are well outside, at the top-left of the edge-on disk. Moreover $PA_{VF}$ is at $\sim$ 50 degrees from the $PA_{opt}$, which definitively discards a rotating disk classification. Both examples illustrate how kinematics and imagery are insufficient when used separately, while combined together, they provide a very powerful technique, the morpho-kinematical classification. If not applied (e.g., without high resolution imagery), one can derive only an upper limit on the rotating disk fraction (see Contini et al., same volume). 

\section{The IMAGES survey}
The explicit goal of IMAGES  is to gather enough constraints on z=0.4-0.8 galaxies for linking them directly to their local descendants. Its selection is limited by an absolute J-band magnitude ($M_{J}(AB)<$ -20.3), a quantity relatively well linked to the stellar mass \citep{Yang08}, leading to a complete sample of 63 emission line galaxies with $W_{0}$([OII])$>$15 A and $M_{stellar}>$ 1.5 $10^{10}$ $M_{\odot}$. The set of measurements includes HST imagery (one to three orbits in b, v, i and z), spatially-resolved kinematics from VLT/GIRAFFE (from 8 to 24hrs integration times), deep VLT/FORS2 observations (two $\times$ three hours with two grisms at R=1500) and deep Spitzer observations to sample mid and near-IR.
IMAGES galaxies have been observed in four different fields of view to avoid cosmological variance effects. It is hardly affected by cosmological dimming, because the imaging depth ensures the detection of the optical disk of a MW-like galaxy after being redshifted to z $\sim$ 0.5. \\

\begin{figure}
\centering
\includegraphics[width=1\columnwidth]{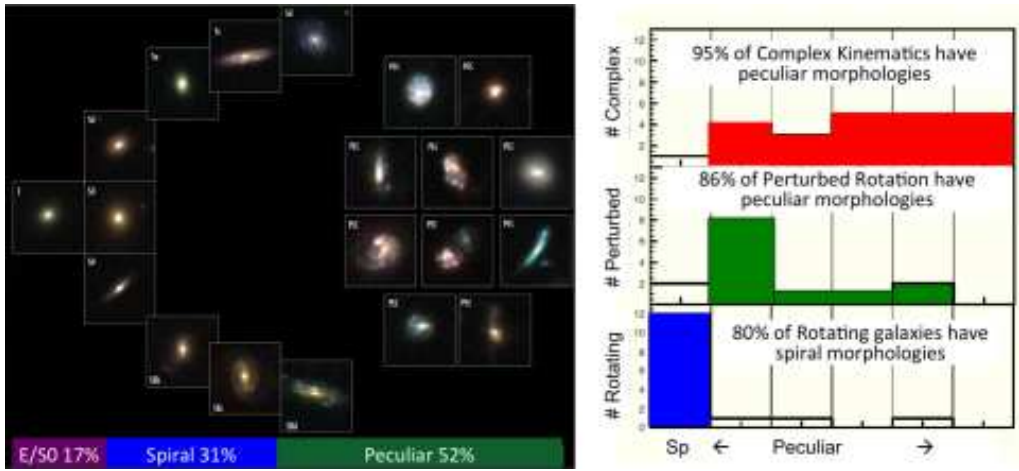} 
\caption{{\it Left}: The past Hubble Sequence, six billion years ago, based on 143 IMAGES galaxies, and for which each stamp represents 5\% of the galaxies.   }
\label{fig:fig2}
\end{figure}

According to the Cosmological Principle, progenitors of present-day giant spirals have properties similar to those of galaxies having emitted their light $\sim$ 6 Gyr ago.
Figure \ref{fig:fig2} presents the results of a morphological analysis of 143 distant galaxies for which depth, rest-frame filters, spatial resolution and selection are strictly equivalent to what has been done for the Hubble Sequence derived from SDSS galaxies \citep{Delgado10}. The past Hubble Sequence shown in Figure \ref{fig:fig2}(left) includes the IMAGES galaxies as well as a complimentary sample of distant galaxies with $W_{0}$([OII])$\le$15 A \citep{Delgado10}. Statistical distributions (stellar mass, star formation, etc.) of both low and high redshift galaxies have been verified to be consistent with the results of large surveys. The methodology for classifying the morphologies follows a semi-automatic decision tree, which uses as templates the well known morphologies of local galaxies that populate the Hubble sequence, including the color of their sub-components. For example a blue nucleated galaxy is classified peculiar. Such a conservative method is the only way for a robust morphological classification, and interestingly, it confirms the \cite{vandenBergh02}'s results.

However star formation may affect the morphological appearance, and the morphological classification has to be compared to the spatially-resolved kinematics. The classification of velocity fields is ranging from rotating disk, perturbed rotation (i.e., $\sigma$ peak offset) or complex kinematics \citep{Flores06,Yang08}.  Figure \ref{fig:fig2} (see also \citealt{Neichel08}) evidences that peculiar morphologies correlate very well with anomalous kinematics and vice versa.

Is that correlation preserved when using automatic classification methods based on e.g., concentration and asymmetry of the galaxy luminosity profile? The answer is negative \citep{Neichel08}, and these methods overestimate the number of spirals by a factor of two, a problem already identified by \cite{Conselice2005}. Automatic classification methods are interesting and appealing because they can be applied to very large number of galaxies. However they provide too simplistic results to interpret complex objects like distant galaxies. Their limitations in distinguishing peculiar from spiral morphologies lead to a far larger systematic than the Poisson statistical uncertainty for a sample of one hundred of objects \citep{Delgado10}, making very large samples treated this way not very useful.


\section{The observed frequency of mergers in the past histories of spirals }
The remarkable agreement between morphological and kinematical classifications implies that dynamical perturbations of the gaseous component at large scales are linked to the peculiar morphological distribution of the stars. This suggests a common process that profoundly modified the overall gas and star properties. 
Minor mergers (mass ratio well above five) can affect locally the dispersion map \citep{Puech07} while they cannot affect the large scale rotational field over several tens of kpc\footnote{This is entirely true for very minor mergers (mass ratio larger than 10) while intermediate mergers may affect significantly the galaxy properties during short phases. Due to their lower impact and their longer duration \citep{Jiang08}, they are considerably
less efficient than major mergers in distorting morphologies and kinematics or they do it in a somewhat sporadic
way \citep{Hopkins08}.}. Outflows provoked by stellar feedback are not observed in the IMAGES galaxies that are generally forming stars at moderate rate \citep{Rodrigues12}. Internal fragmentation is
limited because less than 20\% of the IMAGES sample show clumpy morphologies \citep{Puech10b} while associated cold gas accretion tends
to vanish in massive halos at z$<$1, with rates $<$1.5 $M_{\odot}$/yr at z$\sim$ 0.7 \citep{Keres09}. In fact in many simulations of cold flows, the clumps are merging galaxies containing gas, stars and dark matter (see e.g. Danovich et al. 2011), and for which mergers are easily identified by large jumps in angular momentum acquisition (see e.g., Kimm et al. 2011).
Major mergers appear to be the mechanism explaining peculiar morphologies as well as kinematically unrelaxed galaxies. They are indeed unique in bringing enough angular momentum to explain the large scatter of the mass-velocity (Tully-Fisher) relation \citep{Flores06,Puech08,Covington10}. This explanation indeed solves the spin catastrophe (see also \citealt{Maller02}), i.e. the longstanding problem that disks formed in isolation are too small and with too small angular momentum \citep{Navarro00}.

This has led \cite{Hammer09} to test and then successfully model all the unrelaxed 
IMAGES galaxies as consequences of major mergers  using
hydrodynamical simulations (GADGET2 and ZENO).  Using a comparison of their morpho-kinematics
properties to those from a grid of simple major merger models \citep{Barnes02}, they provided convincing matches in about two-thirds of
the cases. It results that a third of z=0.4-0.8 galaxies are in one or another merger phase. Letting aside the bulge-dominated galaxies (E/S0) which fraction does not evolve (see Figure \ref{fig:fig2}), it means that six billion years ago half of present-day spirals were involved in a major merger phase \citep{Hammer09}. 

Why IMAGES is finding so many spiral progenitors experiencing a major merger phase? In fact the morpho-kinematic technique used in IMAGES is found to be sensitive to all merger phases, from pairs to the post-merger relaxation phase.  The merger rate associated with different phases has been calculated \citep{Puech12}, and found to match perfectly predictions by state-of-the-art $\Lambda$CDM semi-empirical models \citep{Hopkins10} with no particular fine-tuning. Thus, both theory and observations predict an important impact of major mergers for progenitors of present-day spiral galaxies and the whole Hubble sequence could be just a vestige of merger events \citep{Hammer09}. Nowadays all recent cosmological simulations report the formation of late-type disks after major mergers \citep{Font11,Brook11,Keres11,Guedes11, Aumer13}.

\section{Relics of ancient mergers in the nearby Universe}
\begin{figure}
\centering
\includegraphics[width=1\columnwidth]{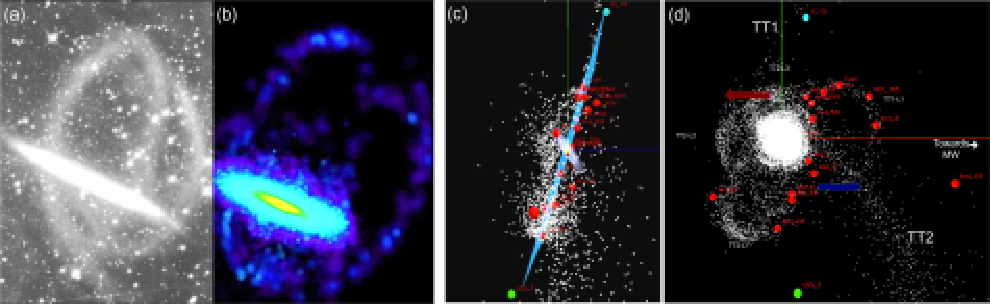} 
\caption{{\it (a)}: Deep observations of NGC 5907 in the visible light showing the thin disk with the gigantic loops (from \citealt{Martinez-Delgado08}). {\it (b)}: GADGET2 simulation (2.2M particles) of the stellar loops, 5.3 Gyrs after a 3:1 merger (first passage, 6.8 Gyrs ago, $f_{gas}^{initial}$=60\%, $f_{baryons}^{initial}$=17\%, see \citealt{Wang12}). {\it (c)}: Stellar particles associated to an ancient 3:1 merger at M31 location (GADGET2, 8M particles see \citealt{Hammer10,Hammer13,Fouquet12}), and forming a plane seen almost edge-on, coinciding with the observed satellite plane (almost seen edge-on, in blue color) found by \cite{Ibata13}. {\it (d)}: Same as (c) but rotated by 90 degrees to reveal the loops system in the loop plane. Most M31's satellite plane (it also includes the recently found CasIII, see \citealt{Martin13}) are found in the loop system and their observed motions follow the predictions from the loop model (see red and blue arrows).
}
\label{fig:fig3}
\end{figure}

If many spiral galaxies have experienced a major merger, it should have left fossil imprints into their halo. Mergers are often producing gigantic tidal tails from which many stars are captured by the halo remnant gravitational potentials \citep{Font06,Wang12,Hammer10,Hammer13}. Because individual stars are not affected by dynamical friction neither by the halo hot gas, they form  stellar streams as illustrated in Figure \ref{fig:fig3}, which can be persistent for a Hubble time.

The outskirts of nearby, isolated spiral galaxies often show low-surface brightness stellar streams \citep{Martinez-Delgado10}, which are likely relics of mergers. An explanation from recent minor mergers seems unlikely because of the stream red colors and of the absence of any residual core. The whole NGC5907 galaxy (disk, bulge and thick disk) and associated loops have been successfully modeled by assuming a 3:1 gas-rich major merger (see Figure \ref{fig:fig3} and \citealt{Wang12}).
Our nearest neighbor, M31 shows a classical bulge and a high halo metallicity suggesting a major merger origin \citep{vandenBergh05,Kormendy13}. 
The later provides a robust explanation of the stellar Giant Stream, which could be made of
tidal tail stars captured by the galaxy gravitational potential after the fusion time \citep{Hammer10}.
Stars of the Giant Stream have ages older than 5.5 Gyr \citep{Brown07}, and a 3:1 gas-rich merger may reproduce it as well as the 
M31 disk, bulge, 10 kpc ring \& thick disk, assuming the interaction and fusion have occurred 8.75$\pm$0.35 and 5.5$\pm$0.5 Gyr ago,
respectively.

In fact, the merger model of M31 of \cite{Hammer10} predicted the disk of satellites discovered by \cite{Ibata13}. Spatial locations of these M31 satellites are drawing two loop systems (see Figure \ref{fig:fig3}d and \citealt{Hammer13}) similar to those found around NGC5907, i.e., typical signatures of ancient major mergers. Besides providing a prediction, the M31 merger model is the only one able to account for the satellite plane thinness, its apparent rotation (see the arrows in Figure \ref{fig:fig3}d) and the excess of dwarves towards the MW (see \citealt{Hammer13} for details). 

Both MW and M31 possess fossil satellite systems, which might be linked together because the \cite{Ibata13} disk of satellites is pointing to the MW. Not surprisingly it is also a prediction of the modeling of M31 as an ancient merger, and the main associated tidal tail may have reached the MW \citep{Fouquet12,Hammer13}. This scenario could be at the origin of the vast plane of satellites observed around the MW \citep{Pawlowski11}, and in such a case many Local Group dSphs would be relics of tidal dwarves, and then without dark matter as verified by \cite{Yang14}.

\section{Discussion \& Conclusions}
The "merger hypothesis" \citep{Toomre72} for elliptical galaxies interprets them as the product of gas-poor or successive major mergers. It appears more and more plausible that spiral galaxies could also result from mergers of gas-rich galaxies \citep{Hammer05,Hammer09}. Re-formation of disks after major mergers has been yet described in 2002 \citep{Barnes02}, assuming a small gas fraction (12\%) in the progenitors\footnote{the rebuilding disk scenario \citep{Hammer05,Hammer09} is also supported by the very recent discovery of rotating molecular disks by ALMA, in nearby merger remnants \citep{Ueda14}.}. With much larger gas fraction a rebuilt disk could be more prominent and after subsequent star formation, it could dominate the remnant galaxy. Gas fractions in galaxies at earliest epochs indicate values that exceed 50\% at z$\sim$ 1.5-2 \citep{Rodrigues12}, allowing even the formation of late type spirals with small B/T values \citep{Hopkins10}.

Perhaps the numerous bulges with low Sersic indices (often called pseudo-bulges, see \citealt{Kormendy13} and references therein) found in local galaxies are at odd with a major merger history. Forming pseudo bulge galaxies is however expected through ancient gas-rich mergers \citep{Keselman12,Hammer12}, though further investigations are necessary. The scatter of the $M_{stellar}-SFR$ relation for emission line galaxies could be too narrow (0.3 dex, \citealt{Noeske07}) for accommodating major mergers, which are expected to show strong and short living bursts during the fusion phase. However the fusion phase is unlikely an instantaneous episode (see e.g., Figure 8 of \citealt{Hopkins12} for mergers of MW-mass galaxies) and they are indeed consistent with duration of 0.4 Gyr for SFR variations by a factor 4 (or 2 $\sigma$). Indeed the shape of the SFR peak depends on the treatment of the overall feedback (see Hopkins, same volume and \citealt{Hopkins12}). Interestingly, observations (from IMAGES, see Figure 4 of \citealt{Puech12}), which average orbits, gas fractions and mass ratios are also fully consistent with the $M_{stellar}-SFR$ scatter \citep{Puech14}.

Because 3D spectroscopy is photon starving, its future is linked to the ELTs.  The combination of the {\em JWST} and of the multi-IFUs MOSAIC on the E-ELT will be uniquely powerful for investigations into the physics underway as the first galaxies were forming (near or at the time when the Universe was only partially reionized).  The potential performances are impressive, i.e. detection of Lyman-$\alpha$ emission line at $z$\,$\sim$\,9 (for $J_{\rm AB}$\,$=$\,29) and the study of rest-frame UV absorption lines in a brighter galaxy at $z$\,$\sim$\,7 ($J_{\rm AB}$\,$=$\,26).  In addition to the larger collecting area of the E-ELT, MOSAIC will provide further gains compared to the first-light instruments envisaged for the GMT and TMT.  Specifically, the use of MOAO-corrected IFUs will provide excellent image quality for spectroscopy over wide fields, enabling the study of spatially-resolved physical properties for large samples of high-redshift galaxies, together with optimum subtraction of the sky background.


\section*{Acknowledgements}

\noindent
We warmly thank both the IMAGES and MOSAIC teams for their support and crucial achievements. We are grateful to the organizers for their invitation at such an excellent Conference and their very nice welcome in the beautiful city of Vienna.

\end{document}